\documentclass[12pt]{article}
\usepackage{amssymb}
\usepackage{amsfonts}
\usepackage{amsmath}
\usepackage[margin=2cm]{geometry}

\def\gz0{\gamma^{0}}

 \def\det{{\rm det\,}}

\def\e{\epsilon}

\def\beq{\begin{equation}}
\newcommand{\eeq}[1]{\label{#1}\end{equation}}
\def\bea{\begin{eqnarray}}
\newcommand{\eea}[1]{\label{#1}\end{eqnarray}}
\def\ba{\begin{array}}
\def\ea{\end{array}}
\def\bec{\begin{center}}
\def\ec{\end{center}}
\def\ba{\begin{align}}
\def\ena{\end{align}}

\def\12{\frac{1}{2}}

\begin{document}

\thispagestyle{empty}

\hfill DFPD/2015/TH-22 ; CERN-PH-TH/2015-232

\vspace{15pt}

\begin{center}

{\Huge Majorana Fermions, Supersymmetry Breaking,\\[3pt]
and Born-Infeld Theory}\\[0pt]


\vspace{35pt} {\Large {S.~Ferrara${}^{\; a,b,c}$, \, A.~Marrani$%
{}^{\; d,e}$}, and A.~Yeranyan${}^{\; b,d}$}\\[20pt]

{${}^a$\textsl{{\ Th-Ph Department, CERN\\[0pt]
CH - 1211 Geneva 23, SWITZERLAND}}} \vspace{10pt}

{${}^b$\textsl{\ INFN - Laboratori Nazionali di Frascati \\[0pt]
Via Enrico Fermi 40, I-00044 Frascati, Italy}}\vspace{10pt}

{${}^c$\textsl{\ Department of Physics and Astronomy, \\[0pt]
U.C.L.A., Los Angeles CA 90095-1547, USA}}\vspace{10pt}

{${}^d$\textsl{\ Centro Studi e Ricerche ``Enrico Fermi'',\\[0pt]
Via Panisperna 89A, I-00184, Roma, Italy}}\vspace{10pt}

{${}^e$\textsl{{\ Dipartimento di Fisica e Astronomia ``Galileo Galilei'',
and INFN,\\[0pt]
Universit\`a di Padova, Via Marzolo 8, I-35131 Padova, Italy
}}}\vspace{10pt}

{${}$\texttt{sergio.ferrara@cern.ch} \\[0pt]
\texttt{alessio.marrani@pd.infn.it}
 \\[0pt]
\texttt{ayeran@lnf.infn.it}}

\vspace{8pt}

\vspace{45pt} \textsc{\large Abstract}
\end{center}

\baselineskip=14pt

\noindent This review is devoted to highlight some aspects of the relevance of
Majorana fermions in rigid supersymmetry breaking in four spacetime
dimensions. After introducing some basic facts on spinors, and on their symmetries and
reality properties, we consider Goldstino actions describing partial
breaking of rigid supersymmetry, then focussing on Born-Infeld non-linear
theory, its duality symmetry, and its supersymmetric extensions, also
including multi-field generalizations exhibiting doubly self-duality.

\vskip 12pt 

\begin{center}
\textsl{\noindent {Contribution to the Proceedings of the Erice
International School of Subnuclear Physics, 53rd Course : ``The Future of
Our Physics Including New Frontiers'', and Celebration of the Triumph of
Ettore Majorana, Erice, 24 June-3 July 2015.}}

\vskip 30pt

\textsl{Dedicated to the memory of Guido Altarelli}
\end{center}

\pagebreak

\noindent \baselineskip=20pt \setcounter{page}{1}

\tableofcontents

\section{Introduction}

The present review is aimed to contribute to the Proceedings of the 53rd
Course of the International School of Subnuclear Physics (held in Erice in
Summer 2015), which also celebrated the Triumph of Ettore Majorana, the
great disappeared Sicilian Physicist whose ideas and invention have
permeated both theoretical and experimental physics.\medskip

Section 2 is devoted to Majorana spinors, their definition in higher
dimensions and their role in Supersymmetry. Example of such particles are
the gravitino and the Goldstino in $D=4$, since the related anticommuting
supersymmetry parameters are Majorana spinors. The same holds true in $%
D=3,9,11$ while in $D=10$ (as in $D=2$) the notion of Majorana-Weyl spinor
occurs. As an application, the role of Majorana and Dirac neutrino masses in
the see-saw mechanism of GUT's is explained.

In Section 3 some basic features of spontaneously broken rigid supersymmetry
and partial supersymmetry breaking are described. The Volkov-Akulov and the
Born-Infeld Lagrangians naturally emerge in this framework, and they both
contribute to the description of partially broken rigid $\mathcal{N}=2$
supersymmetry.

The electric-magnetic self duality enjoyed by the Born-Infeld theory is then
extended in Section 4, in order to describe more general theories, following
the work of Gaillard and Zumino \cite{GZ}. These generalizations involve the Schr\"{o}%
dinger formulation of Born-Infeld-type theories, forms of higher degree as
well as geometrical mass terms in higher dimensions.

\section{Majorana Fermions and Supersymmetry}

Supersymmetry deals with basic spinorial entities called \textit{Majorana
spinors}. This notion originates from a fundamental observation of the
Italian (Sicilian) physicist Ettore Majorana, who noticed that the four-dimensional Dirac equation admits \textquotedblleft real" solutions if the $\gamma $-matrices are suitably chosen \cite{Majorana}.

This is the so-called Majorana representation, in which the $\gamma $%
-matrices are real. Therefore the Dirac equation%
\begin{equation}
\left( \gamma ^{\mu }\partial _{\mu }+m\right) \psi =0  \label{DE}
\end{equation}%
admits manifestly real solutions.

In the Majorana representation, the $\gamma ^{i}$ are symmetric, while $%
\gamma ^{0}$ and $\gamma ^{5}=\gamma ^{0}\gamma ^{1}\gamma ^{2}\gamma ^{3}$
are antisymmetric, and they satisfy the Clifford algebra relations%
\begin{equation}
\left\{ \gamma ^{\mu },\gamma ^{\nu }\right\} =2\eta ^{\mu \nu },~~~~\eta
^{\mu \nu }=diag\left( -,+,+,+\right) .
\end{equation}

How about \textit{chiral} spinors? They cannot be real, since they are
eigenstates of $\gamma ^{5}$, corresponding to the eigenvalues $\pm i$ ($%
\left( \gamma ^{5}\right) ^{2}=-1$) :%
\begin{equation}
\gamma ^{5}\psi _{\pm }=\pm i\psi _{\pm },
\end{equation}%
thus implying that\footnote{%
In the present Section, the upperscript \textquotedblleft $\ast $" denotes
complex conjugation.}
\begin{eqnarray}
\psi _{\pm } &:&=\frac{1}{2}\left( 1\mp i\gamma ^{5}\right) \psi ; \\
\psi _{+}+\psi _{-} &=&\psi ;~~~\psi _{\pm }^{\ast }=\psi _{\mp }.
\end{eqnarray}%
Spinors which are eigenstates of $\gamma ^{5}$ are called \textit{Weyl
spinors}. From the previous properties, it follows that Weyl spinors have
two components instead of four, and they are complex in $D=4$.

Symmetries of $\gamma $-matrices depend on the dimension $D=s+t$ of
spacetime, and the reality properties of spinors on its signature $\rho
:=\left\vert s-t\right\vert $ (\textit{modulo} $8$ : Bott periodicity).
There are eight cases :%
\begin{eqnarray}
\text{complex~spinors} &:&\rho =2,6; \\
\text{real~spinors} &:&\rho =0,1,7; \\
\text{pseudoreal~spinors} &:&\rho =3,4,5,
\end{eqnarray}%
where%
\begin{eqnarray}
\psi ~\text{real} &\Leftrightarrow &\psi ^{\ast }=\psi ; \\
\psi ~\text{pseudoreal} &\Leftrightarrow &\left( \psi _{\alpha }^{A}\right)
^{\ast }=\Omega ^{AB}\mathbb{C}_{\alpha \beta }\psi _{\beta }^{B},
\end{eqnarray}%
with $\Omega $ and $\mathbb{C}$ respectively denoting the anti-involutive
invariant structures of the automorphism group of the supersymmetry algebra
and of the spinor space ($\Omega ^{2}=-1$, $\mathbb{C}^{2}=-1$).

For ordinary $D$-dimensional Minkowski spacetime $\left( s,t\right) =\left(
D-1,1\right) $, and thus $\rho =D-2$. In this case, it holds that%
\begin{eqnarray}
D~\text{even~(\textit{mod}~}8\text{)} &:&\left\{
\begin{array}{l}
\text{complex~spinors}:D=4,8; \\
\text{real~spinors}:D=2,10; \\
\text{pseudoreal~spinors}:D=6,%
\end{array}%
\right. \text{applying~to Weyl spinors~}\psi _{\pm }. \\
&&  \notag \\
D~\text{odd~(\textit{mod}~}8\text{)} &:&\left\{
\begin{array}{l}
\text{real~spinors}:D=3,9,11; \\
\text{pseudoreal~spinors}:D=5,7.%
\end{array}%
\right.
\end{eqnarray}%
For instance, in $D=10$ the Majorana representation is valid for both\ $\psi
_{+}$ and $\psi _{-}$, since $\left( \gamma ^{11}\right) ^{2}=1$; thus, $%
\psi _{+}$ and $\psi _{-}$ in $D=10$-dimensional ordinary Minkowski
spacetime are named \textit{Majorana-Weyl spinors}.

Since we have been mentioning them a few lines above, it is here worth
recalling also the anticommutator sectors of supersymmetry algebras (without
central extensions) in $D$-dimensional Minkowski spacetime:%
\begin{eqnarray}
\left\{ \mathcal{Q}_{\alpha }^{A},\mathcal{Q}_{\beta }^{B}\right\} &=&\Omega
^{AB}\left( \gamma ^{\mu }\right) _{\alpha \beta }P_{\mu }\text{~in~}D=5,6,7;
\label{sa-1} \\
\left\{ \mathcal{Q}_{\alpha }^{A},\mathcal{\bar{Q}}_{\dot{\alpha}B}\right\}
&=&\delta _{B}^{A}\left( \gamma ^{\mu }\right) _{\alpha \dot{\alpha}}P_{\mu }%
\text{~in~}D=4,8;  \label{sa-2} \\
\left\{ \mathcal{Q}_{\alpha }^{A},\mathcal{Q}_{\beta }^{B}\right\} &=&\delta
^{AB}\left( \gamma ^{\mu }\right) _{\alpha \beta }P_{\mu }\text{~in~}%
D=3,9,10,11,  \label{sa-3}
\end{eqnarray}%
where $P_{\mu }$ is the momentum operator. The\ Latin capital indices run
over the number $\mathcal{N}$ of independent spinor charges : $A,B=1,...,%
\mathcal{N}$; for $D$ even, the automorphism group of the supersymmetry
algebra (named $\mathcal{R}$-symmetry group) reads%
\begin{eqnarray}
SO\left( \mathcal{N}_{+}\right) \times SO\left( \mathcal{N}_{-}\right) \text{%
~in~}D &=&2\text{~\textit{mod}~}8~\text{(}\mathcal{N}=\mathcal{N}_{+}+%
\mathcal{N}_{-}\text{)}; \\
Usp\left( 2\mathcal{N}_{+}\right) \times USp\left( 2\mathcal{N}_{-}\right) ~%
\text{in~}D &=&6\text{~\textit{mod}~}8~\text{(}\mathcal{N}=2\mathcal{N}_{+}+2%
\mathcal{N}_{-}\text{)}; \\
U\left( \mathcal{N}\right) \text{~in~}D &=&4,8\text{~\textit{mod}~}8.
\end{eqnarray}

In $D=11$, if the massless spectrum is bound to have $2$ as the highest
spin, then $\mathcal{N}_{\max }=1$, and the corresponding M-theory
superalgebra with central extensions reads as follows :%
\begin{equation}
\left\{ \mathcal{Q}_{\alpha },\mathcal{Q}_{\beta }\right\} =\underset{\text{%
point~particle}}{\left( \gamma ^{\mu }\right) _{\alpha \beta }P_{\mu }}+%
\underset{\text{M2-brane}}{\left( \gamma ^{\mu \nu }\right) _{\alpha \beta
}Z_{\mu \nu }}+\underset{\text{M5-brane}}{\left( \gamma ^{\mu \nu \rho
\sigma \delta }\right) _{\alpha \beta }Z_{\mu \nu \rho \sigma \delta }}.
\end{equation}%
Concerning the counting of (\textit{off-shell}) independent components of
spinors, it goes as follows :%
\begin{eqnarray}
D~\text{even} &:&\text{ each~}\psi _{\pm }\text{ has~}2^{\frac{D-2}{2}}\text{
components~(which double~for~complex or pseudoreal ones)}; \\
D~\text{odd} &:&\text{ each~}\psi \text{ has~}2^{\frac{D-1}{2}}\text{
components~(which double~for pseudoreal ones)}.
\end{eqnarray}%
Consequently, (pseudo)real spinor components of dimension $8$, $16$ and $32$
are possible for $D\leqslant 6$, $D\leqslant 10$ and $D\leqslant 11$,
respectively.

The dimension $D$ \textit{mod }$8$ determines the symmetry of the tensor
products of spinor representations (denoted by $\psi $), which are defined
by the \textit{morphism} map%
\begin{equation}
A:\psi \times \psi \rightarrow \Lambda ^{k},
\end{equation}%
where $\Lambda ^{k}$ denotes the space of $k$-forms. The resulting symmetry
of $A$, depending on $D$ and $k$, is presented in Table 1.

\begin{table}[th]
\begin{center}
\begin{tabular}{|c|c|c||c|c|}
\hline
\multicolumn{1}{|c|}{$D~($mod~8$)$} & \multicolumn{2}{|c||}{$k$ even} &
\multicolumn{2}{|c|}{$k$ odd} \\ \hline\hline
& morphism & symmetry & morphism & symmetry \\ \hline
0 & $\psi _{\pm }\otimes \psi _{\pm }\rightarrow \Lambda ^{k}$ & $%
(-1)^{k(k-1)/2}$ & $\psi _{\pm }\otimes \psi _{\mp }\rightarrow \Lambda ^{k}$
&  \\ \hline
1 & $\psi \otimes \psi \rightarrow \Lambda ^{k}$ & $(-1)^{k(k-1)/2}$ & $\psi
\otimes \psi \rightarrow \Lambda ^{k}$ & $(-1)^{k(k-1)/2}$ \\ \hline
2 & $\psi _{\pm }\otimes \psi _{\mp }\rightarrow \Lambda ^{k}$ &  & $\psi
_{\pm }\otimes \psi _{\pm }\rightarrow \Lambda ^{k}$ & $(-1)^{k(k-1)/2}$ \\
\hline
3 & $\psi \otimes \psi \rightarrow \Lambda ^{k}$ & $-(-1)^{k(k-1)/2}$ & $%
\psi \otimes \psi \rightarrow \Lambda ^{k}$ & $(-1)^{k(k-1)/2}$ \\ \hline
4 & $\psi _{\pm }\otimes \psi _{\pm }\rightarrow \Lambda ^{k}$ & $%
-(-1)^{k(k-1)/2}$ & $\psi _{\pm }\otimes \psi _{\mp }\rightarrow \Lambda
^{k} $ &  \\ \hline
5 & $\psi \otimes \psi \rightarrow \Lambda ^{k}$ & $-(-1)^{k(k-1)/2}$ & $%
\psi \otimes \psi \rightarrow \Lambda ^{k}$ & $-(-1)^{k(k-1)/2}$ \\ \hline
6 & $\psi _{\pm }\otimes \psi _{\mp }\rightarrow \Lambda ^{k}$ &  & $\psi
_{\pm }\otimes \psi _{\pm }\rightarrow \Lambda ^{k}$ & $-(-1)^{k(k-1)/2}$ \\
\hline
7 & $\psi \otimes \psi \rightarrow \Lambda ^{k}$ & $(-1)^{k(k-1)/2}$ & $\psi
\otimes \psi \rightarrow \Lambda ^{k}$ & $-(-1)^{k(k-1)/2}$ \\ \hline
\end{tabular}%
\end{center}
\caption{Properties of the morphism map $A$ \protect\cite{Spinor-Algebras}}
\label{morphisms}
\end{table}
For a comprehensive treatment of spinors and Clifford algebras for arbitrary
$\left( s,t\right) $-signatures of spacetime, see \textit{e.g.} \cite%
{Spinor-Algebras}.

It should be recalled that the massless Dirac equation ((\ref{DE}) with $m=0$%
) has an extra symmetry, which in the Majorana representation takes the form%
\begin{equation}
\psi ^{\prime }=e^{\alpha \gamma ^{5}}\psi ;
\end{equation}%
this is indeed a $U(1)$ rotation, since $\gamma ^{5}$ is anti-involutive : $%
\left( \gamma ^{5}\right) ^{2}=-1$.

However, this symmetry is broken by the mass term for a Majorana fermion
(called \textit{Majorana mass}). Defining a complex Dirac spinor $\psi _{D}$
as%
\begin{eqnarray}
\psi _{D} &:&=\psi _{1}+i\psi _{2}; \\
\psi _{D}^{\ast } &=&\psi _{1}-i\psi _{2},
\end{eqnarray}%
where $\psi _{1}$ and $\psi _{2}$ are Majorana spinors, one can them have a $%
U(1)$ symmetry which rotates $\psi _{1}$ into $\psi _{2}$.

Then, one can define%
\begin{eqnarray}
\psi _{L} &:&=\frac{1}{2}\left( 1-i\gamma ^{5}\right) \psi _{D};~~~\psi
_{R}:=\psi _{L}^{\ast }=\frac{1}{2}\left( 1+i\gamma ^{5}\right) \psi
_{D}^{\ast }; \\
\chi _{L} &:&=\frac{1}{2}\left( 1-i\gamma ^{5}\right) \psi _{D}^{\ast
};~~~\chi _{R}:=\chi _{L}^{\ast }=\frac{1}{2}\left( 1+i\gamma ^{5}\right)
\psi _{D}.
\end{eqnarray}%
Therefore, $\psi _{L}$ and $\chi _{L}$ (and $\psi _{R}$ and $\chi _{R}$)
have opposite $U(1)$ phases, whereas $\psi _{L}$ and $\chi _{R}$ (and $\psi
_{R}$ and $\chi _{L}$) have identical $U(1)$ phases. The $U(1)$-invariant
Dirac equation in the chiral (Weyl) notation becomes%
\begin{equation}
\gamma ^{\mu }\partial _{\mu }\psi _{L}+m\chi _{R}=0~\left( +h.c.\right)
;~~~\gamma ^{\mu }\partial _{\mu }\chi _{L}+m\psi _{R}=0~\left( +h.c.\right)
.
\end{equation}%
Thus, the Dirac mass term, in this notation, is $m\psi _{L}\chi _{L}+h.c.$,
which is of course $U(1)$-invariant.

In principle, if one gives up the $U(1)$ symmetry, \textit{a Dirac fermion
can have three types of mass terms}, namely :%
\begin{equation}
m\psi _{L}\chi _{L},~~~~M\psi _{L}\psi _{L},~~~~N\chi _{L}\chi _{L},
\end{equation}%
where $m$ is named \textit{Dirac mass}, and $M$ and $N$ \textit{Majorana
masses}. As a result, the matrix%
\begin{equation}
\mathcal{M}:=\left(
\begin{array}{cc}
M & m \\
m & N%
\end{array}%
\right)
\end{equation}%
has two eigenvalues :%
\begin{equation}
m_{1,2}:=\frac{1}{2}\left[ M+N\pm \sqrt{\left( M-N\right) ^{2}+4m^{2}}\right]
.
\end{equation}%
In particular, for $N=0$ and $M\gg m$, the eigenvalues become%
\begin{equation}
m_{1,2}\cong \left( M,-\frac{m^{2}}{M}\right) .
\end{equation}%
This result is at the basis of the \textit{see-saw mechanism} \cite%
{Minkowski,GRS}.

Concerning the superpartner of the graviton, this is a spin-$\frac{3}{2}$
Rarita-Schwinger field, called \textit{gravitino}. Under the Lorentz group $%
SO(1,D-1)$ it is a vector-spinor $\psi _{\mu \alpha }$. If it is massless
(namely, in the case of unbroken supersymmetry), it lies in an irreducible, $%
\gamma $-traceless representation of the little group $SO(D-2)$; so,
assuming real (semi)spinors, the gravitino has a number of (\textit{on-shell}%
) independent components given by%
\begin{eqnarray}
D~\text{odd, (}\psi ~\text{real)} &:&\#\left( \psi _{\mu \alpha }\right)
=\left( D-2\right) 2^{\frac{D-3}{2}}-2^{\frac{D-3}{2}}=\left( D-3\right) 2^{%
\frac{D-3}{2}}; \\
D~\text{even, (}\psi _{\pm }~\text{real)} &:&\#\left( \psi _{\mu \alpha ,\pm
}\right) =\left( D-2\right) 2^{\frac{D-4}{2}}-2^{\frac{D-4}{2}}=\left(
D-3\right) 2^{\frac{D-4}{2}},
\end{eqnarray}%
which for example yields%
\begin{eqnarray}
D &=&11:\#\left( \psi _{\mu \alpha }\right) =8\cdot 2^{4}=128; \\
D &=&10:\#\left( \psi _{\mu \alpha ,\pm }\right) =7\cdot 2^{3}=56.
\end{eqnarray}%
The simplest Lagrangian density of supergravity couples Einstein gravity
with the Rarita-Schwinger density, yielding%
\begin{equation}
\mathcal{L}_{\text{sugra}}=R-\bar{\psi}_{\mu }\gamma ^{\mu \nu \rho }D_{\nu
}\psi _{\rho }.
\end{equation}%
Assuming 2 as the highest spin in the massless spectrum (pertaining to the
graviton), after \cite{Nahm} it is known that Poincar\'{e} supersymmetry
coupled to gravity is possible for $D\leqslant 11$, while conformal
supersymmetry is possible for $D\leqslant 6$; this result also stems from
the classification of superalgebras done by Kac \cite{Kac}.

\section{Spontaneous Symmetry Breaking of Rigid Supersymmetry : Goldstino
Actions}

As holding for any action invariant under continuous symmetries, in theories
invariant under one or more supersymmetry transformations there are
conserved \textit{Majorana vector--spinor Noether currents} \cite{fz} (in
van der Waerden notation):%
\begin{equation}
\partial ^{\mu }\,\ J_{\mu \,\alpha }^{A}(x)\ =\ 0\quad (A=1,..,\mathcal{N}).
\end{equation}
The corresponding charges
\begin{equation}
\mathcal{Q}_{\alpha }^{A}\ =\ \int d^{3}\mathbf{x}\,\ J_{0\,\alpha }^{A}(x)
\end{equation}
generate the ($\mathcal{N}$--extended) supersymmetry algebra, whose
non-trivial part in $D=4$ (which we consider throughout this Section) is
given by (\ref{sa-2}).

In $\mathcal{N}=1$ spontaneous breaking, the order parameter $\mu$ enters a
term linear in the supercurrent \cite{salam,FI}%
\begin{equation}
J_{\mu \,\dot{\alpha}}\ =\ \mu\,\ \left( \gamma _{\mu }{\mathcal G}\right) _{\dot{\alpha}%
}+\ ...\ ,
\end{equation}
where $\mathcal{G}$ is a Majorana field, the \textit{Goldstino; note that} $\mu$ has
dimension $2$ in natural units.

The supersymmetry current algebra implies that%
\begin{equation}
\int d^{3}\mathbf{y}\left\{ J_{0\,\dot{\alpha}}(y),J_{\mu \alpha }\left(
x\right) \right\} =\left( \sigma ^{\nu }\right) _{\alpha \dot{\alpha}}T_{\nu
\mu }\left( x\right) +\text{derivatives},
\end{equation}%
so that%
\begin{equation}
\langle 0|\left\{ \mathcal{\bar{Q}}_{\dot{\alpha}\,},J_{\mu \alpha
}(x)\right\} |0\rangle \ =\ \left( \sigma ^{\nu }\right) _{\alpha \,\dot{%
\alpha}}\ \langle 0|T_{\nu \mu }|0\rangle \,\ =\ \left( \sigma _{\mu
}\right) _{\alpha \,\dot{\alpha}}\ \mu^{2}\,\ .
\end{equation}%
By generalizing this for several supercharges, one can conclude that (in
absence of central extensions, as well as in the rigid case - no gravity coupled -, as we assume throughout this Section),
either all supersymmetries are unbroken or they are all broken at the same
scale $f$ :%
\begin{equation}
J_{\mu \,\dot{\alpha}A}\ =\ \mu\,\ \left( \gamma _{\mu }{\mathcal G}_{A}\right) _{\dot{%
\alpha}}+\ ...\ .
\end{equation}

One can evade this result by modifying the current algebra via terms not
proportional to $\delta _{A}^{B}$, namely by adding a contribution
proportional to a constant matrix $C_{A}^{B}$ in the adjoint of $SU(\mathcal{%
N})$ (\textit{i.e.}, traceless and Hermitian):%
\begin{equation}
\langle 0|\left\{ \mathcal{\bar{Q}}_{\dot{\alpha}\,},J_{\mu \alpha
}(x)\right\} |0\rangle \ =\ \left( \sigma ^{\nu }\right) _{\alpha \,\dot{%
\alpha}}\ \langle 0|T_{\nu \mu }|0\rangle \,\ +\left( \sigma _{\mu }\right)
_{\alpha \,\dot{\alpha}}C_{A}^{B}.\label{44}
\end{equation}%
As it holds in general for Goldstone particles, the Goldstino $\mathcal G$ is a
massless fermion, with low--energy self--interactions described by a
Lagrangian invariant under a non--linear realization of supersymmetry, and
depending on both $\mathcal{N}$ and $C_{A}^{B}$, because it can be shown that%
\begin{equation}
\delta _{A}\chi ^{i}\,\ \delta ^{B}{\overline{\chi }}_{i}\ =\ V\,\ \delta
_{A}^{B}\ +\ C_{A}^{B}\quad \left( \delta \chi ^{i}\ =\ \delta _{A}\chi
^{i}\,\epsilon ^{A}\right) \ .
\end{equation}%
If $k$ supersymmetries are unbroken, the Hermitian matrix $V\mathbf{{1}+C}$
has rank $\mathcal{N}-k$ in the vacuum, and the $\mathcal{N}-1$ possible
scales of partial supersymmetry breaking are the classified by the $\mathcal{%
N}-1$ Casimirs of $SU(\mathcal{N})$. Amusingly, the resulting characteristic
equation can be solved in algebraically closed form up to the quartic order,
corresponding in this setting to $\mathcal{N}=4$, the maximal value of $%
\mathcal{N}$ for rigid supersymmetry.

For $\mathcal{N}=2$, there are various examples realizing a spontaneous
supersymmetry breaking \cite{BG-1,BG-2,Pol,Ferrara,Anto}, and the previous analysis in
terms of $SU(2)$ invariants was performed in \cite{adft_15}. The rigid limit of $\mathcal{N}=2$ supergravity theory reproducing partial
supersymmetry breaking was considered in \cite{ACDRT}. The unique
Casimir of $SU(2)$ is the squared norm of a 3--vector $\xi ^{x}$ constructed
via electric and magnetic Fayet--Iliopoulos (FI) terms%
\begin{eqnarray}
\xi ^{x} &:&=\left( Q_{y}\wedge Q_{z}\right) \epsilon ^{xyz}; \\
Q_{y}\wedge Q_{z} &:&=m_{y}^{\Lambda }\,\ e_{z\Lambda }\ -\ m_{z}^{\Lambda
}\,\ e_{y\Lambda }\ ,
\end{eqnarray}%
where $Q_{x}=(m_{x}^{\Lambda },e_{x\,\Lambda })$, and $\Lambda =1,\ldots ,n$%
, with $n$ denoting the number of $\mathcal{N}=2$ vector multiplets. The
wedge product $\wedge $ is with respect to the symplectic structure
inherited from the rigid special K\"{a}hler geometry endowing the vector
multiplets' scalar manifold, and the matrix $C_{A}^{B}$ acquires the
explicit form%
\begin{equation}
\delta _{A}\chi ^{i}\,\ \delta ^{B}\overline{\chi }_{i}\ =\ V\,\ \delta
_{A}^{B}\ +\ \left( \sigma _{x}\right) _{A}^{B}\,\ \xi ^{x}\ =\ \left(
\begin{array}{cc}
V-\xi ^{3} & \xi ^{1}+i\,\xi ^{2} \\
\xi ^{1}-i\,\xi ^{2} & V+\xi ^{3}%
\end{array}%
\right) ,
\end{equation}%
with eigenvalues%
\begin{equation}
\lambda _{\pm }=V\mp \sqrt{\xi ^{x}\xi ^{x}}.
\end{equation}%
Thus, when $V=\sqrt{\xi ^{x}\xi ^{x}}$, $\lambda _{+}=0$ and $\mathcal{N}=2$
rigid supersymmetry is broken down to $\mathcal{N}=1$, with order parameter%
\begin{equation}
\mu=\left( \xi ^{x}\xi ^{x}\right) ^{1/4}.
\label{op}
\end{equation}

The last term in the r.h.s. of Eq. (\ref{44}) corresponds to a
\textquotedblleft vector" central charge $Z_{\mu \mid B}^{A}$ \cite{DS,FP};
the only field not being inert is the gauge field of $\mathcal{N}=2$ vector
multiplets.

For a general $\mathcal{N}$-extended rigid supersymmetry broken down to $%
\mathcal{N}=0$ (corresponding, from the previous reasoning, to $C_{A}^{B}=0$%
), the low--energy effective Lagrangian for $N$ Goldstino fields is an
obvious extension of the Volkov--Akulov (VA) action \cite{va} derived in the
$\mathcal{N}=1$ case, and in general reads%
\begin{equation}
\mathcal{L}_{VA}\left( \mu,{\mathcal G}_{\alpha }^{A}\right) \ =\ \mu^{2}\left[ 1\ -\ \sqrt{%
\,-\,\mathrm{det}\left( \eta _{\mu \nu }\ +\ \frac{i}{\mu^{2}}\ \left(
\overline{{\mathcal G}}_{A}\gamma _{\mu }\partial _{\nu }\,{\mathcal G}^{A}-h.c.\right) \right) }%
\right] \ .
\end{equation}%
It should be stressed that this Lagrangian contains a finite number of
terms, because of the \textit{nilpotency} of the Goldstino fields ${\mathcal G}^{A}$.
Furthermore, $\mathcal{L}_{VA}$ has a $U(\mathcal{N})$ $\mathcal{R}$%
--symmetry, being invariant under the following non--linear supersymmetry
transformations :%
\begin{equation}
\delta \,\ {\mathcal G}_{\alpha }^{A}(x)\ =\ \mu \,\ \epsilon _{\alpha }^{A}\ +\ \frac{i}{\mu%
}\,\ \left( \overline{{\mathcal G}}_{B}\,\gamma ^{\mu }\epsilon ^{B}\ -\ \overline{%
\epsilon }_{B}\,\gamma ^{\mu }{\mathcal G}^{B}\right) \ \partial _{\mu }\,\ {\mathcal G}_{\alpha
}^{A}\ .
\end{equation}

On the other hand, the Goldstino action for $\mathcal{N}=2$ rigid
supersymmetry partially broken down to $\mathcal{N}=1$ is described by the
supersymmetric Born--Infeld (SBI) theory : this is given by a \textit{%
non-linear} Lagrangian $\mathcal{L}_{SBI}\left( {\mathcal G}_{\alpha },F_{\mu \nu
}\right) $ for an $\mathcal{N}=1$ vector multiplet whose chiral superfield
strength $W_{\alpha }=\overline{D}^{\,2}\,D_{\alpha }\,V$ contains the
goldstino ${\mathcal G}$ at $\theta =0$ and the self--dual Maxwell field strength at
the next order in $\theta $, with the following two properties :
\begin{eqnarray}
&&\mathcal{L}_{SBI}\left( {\mathcal G}_{\alpha },F_{\mu \nu }=0\right) \
\longrightarrow \ \mathcal{L}_{VA}\ ; \\
&&\mathcal{L}_{SBI}\left( {\mathcal G}_{\alpha }=0,F_{\mu \nu }\right) \ =\ \mu^{2}\left[
1\ -\ \sqrt{\ -\ det\left( \eta _{\mu \nu }\ +\ \frac{1}{\mu}\ F_{\mu \nu
}\right) }\right] \ =\mathcal{L}_{BI},  \label{susyb13}
\end{eqnarray}%
where $\mathcal{L}_{BI}$ is given by (\ref{BI-1}) further below.

$\mathcal{L}_{SBI}$ exhibits two types of super--invariances: a manifest $%
\mathcal{N}=1$ supersymmetry, linearly realized in $\mathcal{N}=1$
superspace, and a second supersymmetry which is non--manifest and
non--linearly realized. In terms of microscopic parameters, the scale of the
broken supersymmetry is the $SU(2)$- and symplectic- invariant quantity (\ref%
{op}); see Sec. \ref{SUSY-SKG} further below.

It is here worth pointing out that the Goldstino action for a partial
breaking $\mathcal{N}\rightarrow \mathcal{N}-k$ of rigid supersymmetry is
only known for $\mathcal{N}=2$, corresponding to the non--linear limit of a
quadratic action of $\mathcal{N}=2$ vector multiplets endowed with FI terms. The simplest case is given by a single $N=2$ vector
multiplet, which under the spontaneous breaking $\mathcal{N}=2\rightarrow 1$
reduces to an $\mathcal{N}=1$ vector multiplet $W_{\alpha }$ and to an $%
\mathcal{N}=1$ chiral multiplet $X$. The presence of a non-vanishing matrix $%
C_{A}^{B}$ allows the chiral multiplet $X$ to acquire a non-zero mass $m_{X}$
:%
\begin{equation}
\underset{\mathcal{N}=2\text{ vector mult.}}{\left( 1,2\left( \frac{1}{2}%
\right) ,2\left( 0\right) \right) }\ \overset{\mathcal{N}=2\rightarrow 1}{%
\longrightarrow }\ \underset{\mathcal{N}=1\text{ vector mult., }m_{V}=0}{%
\left( 1,\frac{1}{2}\right) }\ +\ \underset{\mathcal{N}=1\text{ chiral
mult., }m_{X}\neq 0}{\left( \frac{1}{2},2\left( 0\right) \right) }\ .
\end{equation}%
For $m_{X}$ large enough, $X$ can be integrated out, giving rise to a
non--linear theory for the fields ${\mathcal G}_{\alpha }$ and $F_{\mu \nu }$ of the $%
\mathcal{N}=1$ vector multiplet, described by the aforementioned $\mathcal{L}%
_{SBI}\left( {\mathcal G}_{\alpha },F_{\mu \nu }\right) $ \cite{fps,fpssy}.

\section{Electric-Magnetic Duality, Born--Infeld and Generalizations}

\textit{Electric-magnetic (e.m.) duality} is one of the most fascinating
symmetries of non-linear (and thus interacting) theories of gauge fields.

\subsection{Maxwell Theory}

\textit{Free Maxwell theory is the prototype of e.m. duality invariant
theory.} In the vacuum, the Eqs. of motion read ($F_{\mu \nu }:=\partial
_{\mu }A_{\nu }-\partial _{\nu }A_{\mu }$)
\begin{equation}
\partial _{\mu }F^{\mu \nu }=0,~~\partial _{\mu }~\star F^{\mu \nu }=0
\end{equation}%
where
\begin{equation}
\star F_{\mu \nu }:=\frac{1}{2}\epsilon _{\mu \nu \rho \sigma }F^{\rho
\sigma }~,
\end{equation}%
and they are invariant under $SO(2)$ rotations:%
\begin{equation}
\left(
\begin{array}{c}
F \\
\star F%
\end{array}%
\right) ^{\prime }=\left(
\begin{array}{cc}
\cos \alpha & -\sin \alpha \\
\sin \alpha & \cos \alpha%
\end{array}%
\right) \left(
\begin{array}{c}
F \\
\star F%
\end{array}%
\right) .  \label{ff'}
\end{equation}%
This latter transformation, in terms of electric and magnetic fields $F_{\mu
\nu }=\left( \overrightarrow{E},\overrightarrow{B}\right) $ reads
\begin{equation}
\left(
\begin{array}{c}
\overrightarrow{E} \\
\overrightarrow{B}%
\end{array}%
\right) ^{\prime }=\left(
\begin{array}{cc}
\cos \alpha & -\sin \alpha \\
\sin \alpha & \cos \alpha%
\end{array}%
\right) \left(
\begin{array}{c}
\overrightarrow{E} \\
\overrightarrow{B}%
\end{array}%
\right) .
\end{equation}%
Also in this notation it is immediate to see that Maxwell Equations in
absence of sources ($F_{oi}=E_{i}$, $F_{ij}=\epsilon _{ijk}B_{k}$)%
\begin{eqnarray}
\overrightarrow{\nabla }\cdot \overrightarrow{E} &=&0,~~~\partial _{t}%
\overrightarrow{E}=\overrightarrow{\nabla }\times \overrightarrow{B}, \\
\overrightarrow{\nabla }\cdot \overrightarrow{B} &=&0,~~~\partial _{t}%
\overrightarrow{B}=-\overrightarrow{\nabla }\times \overrightarrow{E}
\end{eqnarray}%
are invariant.

Note that the Hamiltonian%
\begin{equation}
\mathcal{H}=\frac{1}{2}\left( \left\vert \overrightarrow{E}\right\vert
^{2}+\left\vert \overrightarrow{B}\right\vert ^{2}\right)
\end{equation}%
is e.m. duality invariant, while the Lagrangian

\begin{equation}
\mathcal{L}=\frac{1}{2}\left( \left\vert \overrightarrow{E}\right\vert
^{2}-\left\vert \overrightarrow{B}\right\vert ^{2}\right)
\end{equation}%
is not. It should also be pointed out that the \textit{e.m. duality
symmetry\ is not an internal symmetry, since it rotates tensors with
pseudo-tensors} (as such, it can be regarded as a sort of \textquotedblleft
bosonic chiral transformation").\newline

Electric-magnetic dualities are transformations among two-form field
strengths and their \textit{duals}. These transformations extend to $(p+2)$%
-form field strengths and their duals, which, in $D$ space-time dimensions,
are $(D-p-2)$-forms (for $D=4$ and $p=0$, one obtains that $p+2$ $=D-p-2$).
In order to generalize the e.m. duality to $p$-forms, one demands that $%
D/2=p+2$, so that one has $(p+1)$-form gauge fields coupled to sources,
which are $p$-extended objects: $p=0$ (pointlike), $p=1$ (string), $p=2$
(membrane), $p=3$ (three-brane), \textit{etc.}:%
\begin{equation}
F_{\mu _{1}...\mu _{p+2}}=\partial _{\lbrack \mu _{1}}A_{\mu _{2}...\mu
_{p+2}]}.
\end{equation}

When $D/2=p+2$ we have that $D-p-2=p+2$, and $p$-dimensional extended
objects can source both electric and magnetic fields. We thus have \textit{%
dyons}, extended objects that carry both an electric and a magnetic charge.
In all other cases, electric and magnetic objects extends in different space
dimensions, ($p$, $D-p-4$) respectively, so dyons \textit{cannot} exist.

The Dirac-Schwinger-Zwanziger quantization condition reads as follows:

\begin{itemize}
\item for $D/2\neq p+2$:%
\begin{equation}
em^{\prime }=2\pi k,~~~k\in \mathbb{Z},
\end{equation}

\item for $D/2=p+2$
\begin{eqnarray}
em^{\prime }+e^{\prime }m &=&2k\pi ,~~k\in \mathbb{Z}\text{~(}p~\text{odd)};
\\
em^{\prime }-e^{\prime }m &=&2k\pi ,~~k\in \mathbb{Z}\text{~(}p~\text{even)}.
\end{eqnarray}
\end{itemize}

Note that the latter condition is not only invariant under $SO(2)$, but also
under $Sp(2,\mathbb{R})$, isomorphic to $SL(2,\mathbb{R})$.

\subsection{Non-Linear Electric--Magnetic Duality}

The very idea of duality is to generalize the duality of free Maxwell theory
to the case of interactions. Interactions introduce non-linearities, which
can be due either to the Maxwell field itself, or to other matter fields (as
in electrodynamics) which interact with the e.m. field.


\textit{Non-linear} (pure) electromagnetism can be seen as an effective
theory of Maxwell theory in a medium that has a nonlinear response to the
electric and magnetic fields. These nonlinearities are captured by the
relations expressing the \textit{electric displacement} $\overrightarrow{D}$
and the \textit{magnetic field} $\overrightarrow{H}$ in terms of the \textit{%
magnetic induction} $\overrightarrow{B}$ and the \textit{electric field} $%
\overrightarrow{E}$. The field equations (in absence of sources) are thus
divided in a set of linear ones that are independent from the medium
\begin{eqnarray}
\overrightarrow{\nabla }\cdot \overrightarrow{D} &=&0,~~~\partial _{t}%
\overrightarrow{D}=\overrightarrow{\nabla }\times \overrightarrow{H},
\label{eomlinear} \\
\overrightarrow{\nabla }\cdot \overrightarrow{B} &=&0,~~~\partial _{t}%
\overrightarrow{B}=-\overrightarrow{\nabla }\times \overrightarrow{E},
\end{eqnarray}%
%
%
%
%
%
%
and a set describing the nonlinearities, i.e. the properties of the medium,
hence these equations are called constitutive relations,
\begin{eqnarray}
\overrightarrow{D} &=&\overrightarrow{D}\left( \overrightarrow{E},%
\overrightarrow{B}\right) =\overrightarrow{E}+...,  \label{nonlinearDE0} \\
\overrightarrow{H} &=&\overrightarrow{H}\left( \overrightarrow{E},%
\overrightarrow{B}\right) =\overrightarrow{B}+...,  \label{nonlinearDE}
\end{eqnarray}%
Maxwell theory is recovered when
\begin{equation}
\overrightarrow{D}=\overrightarrow{E},~~~\overrightarrow{H}=\overrightarrow{B%
}~.  \label{linear}
\end{equation}%
The dots $\ldots $ in Eq. (\ref{nonlinearDE}) stand for some higher powers
of the electric field $\overrightarrow{E}$ and of the magnetic induction $%
\overrightarrow{B}$, so that the nonlinear theories we consider are
deformations of linear electromagnetism because (\ref{nonlinearDE0}) and (%
\ref{nonlinearDE}) reduce to (\ref{linear}) for weak fields.

In a theory based on a Lagrangian density $\mathcal{L}=\mathcal{L}\left(
\overrightarrow{E},\overrightarrow{B}\right) $, one has the Eqs.
\begin{equation}
\overrightarrow{D}=2\frac{\delta \mathcal{L}\left( \overrightarrow{E},%
\overrightarrow{B}\right) }{\delta \overrightarrow{E}},~~~\overrightarrow{H}%
=-2\frac{\delta \mathcal{L}\left( \overrightarrow{E},\overrightarrow{B}%
\right) }{\delta \overrightarrow{B}}.
\end{equation}

Note that Eqs. (\ref{eomlinear}) are invariant under the linear
transformations
\begin{eqnarray}  \label{BDrot}
\left(
\begin{array}{c}
\overrightarrow{B} \\
-\overrightarrow{D}%
\end{array}%
\right) ^{\prime } &=&\left(
\begin{array}{cc}
A & B \\
C & D%
\end{array}%
\right) \left(
\begin{array}{c}
\overrightarrow{B} \\
-\overrightarrow{D}%
\end{array}%
\right) ; \\
\left(
\begin{array}{c}
\overrightarrow{E} \\
\overrightarrow{H}%
\end{array}%
\right) ^{\prime } &=&\left(
\begin{array}{cc}
A & B \\
C & D%
\end{array}%
\right) \left(
\begin{array}{c}
\overrightarrow{E} \\
\overrightarrow{H}%
\end{array}%
\right) .  \label{EHrot}
\end{eqnarray}
Hence the full set of equations of motion are invariant under the $\left(%
\begin{array}{cc}
A & B \\
C & D%
\end{array}%
\right)$ duality rotations if the constitutive relations are compatible with
these rotations, i.e. if the transformations $\overrightarrow{D}(%
\overrightarrow{E}, \overrightarrow{B}) \to\overrightarrow{D}(%
\overrightarrow{E^{\prime }}, \overrightarrow{B^{\prime }})$ and $%
\overrightarrow{H}(\overrightarrow{E}, \overrightarrow{B}) \to%
\overrightarrow{H}(\overrightarrow{E^{\prime }}, \overrightarrow{B^{\prime }}%
)$ are the same as the rotations (\ref{BDrot}), (\ref{EHrot}).\newline

It is convenient to rewrite the nonlinear equations and the duality
rotations using a relativistic formalism. We set as usual $F_{\mu \nu
}=\left( \overrightarrow{E},\overrightarrow{B}\right) $ and $G_{\mu \nu
}=\left( \overrightarrow{H},-\overrightarrow{D}\right) $, so that in the
vacuum:%
\begin{equation}
\overrightarrow{D}=\overrightarrow{E},~~~\overrightarrow{H}=\overrightarrow{B%
}~~\Leftrightarrow ~G_{\mu \nu }=\star F_{\mu \nu }.
\end{equation}%
The dynamical equations then read $dF=0,dG=0$ i.e.,
\begin{equation}
\partial _{\mu \,}\star {F}^{\mu \nu }=0,~~~\partial _{\mu \,}\star {G}^{\mu
\nu }=0~,
\end{equation}%
while the constitutive relations become
\begin{eqnarray}
G_{\mu \nu } &=&G_{\mu \nu }\left( F,\,\star F\right) , \\
G_{\mu \nu } &=&\star F_{\mu \nu }+...~,
\end{eqnarray}%
and, if we have a Lagrangian,
\begin{equation}
\star G^{\mu \nu }=2\frac{\delta \mathcal{L}\left( F\right) }{\delta F_{\mu
\nu }}.  \label{orig-rels}
\end{equation}%
In order to have compatibility with the equations of motion, this demands
the following integrability conditions :%
\begin{equation}
\frac{\delta \star G^{\mu \nu }}{\delta F^{\rho \sigma }}=\frac{\delta \star
G_{\rho \sigma }}{\delta F_{\mu \nu }}.
\end{equation}

The duality rotations correspond to transform%
\begin{equation}
\left(
\begin{array}{c}
F \\
G%
\end{array}%
\right) \rightarrow \left(
\begin{array}{c}
F \\
G%
\end{array}%
\right) ^{\prime }=\left(
\begin{array}{cc}
A & B \\
C & D%
\end{array}%
\right) \left(
\begin{array}{c}
F \\
G%
\end{array}%
\right) ,  \label{lin-transf}
\end{equation}%
and are a symmetry of the theory if the transformation $G(F)\rightarrow
G(F^{\prime })=G(AF+BG)$ is the same as the rotation $G(F)\rightarrow
CF+DG(F)$. This highly non-linear constraint restricts the possible forms of
$G$ (the constitutive relations), and also the linear transformations (\ref%
{lin-transf}). \newline

Gaillard and Zumino \cite{GZ} have proven that the most general e.m. duality
rotation for a non-linear theory, also depending on extra matter fields
(such as fermions, scalars, and possibly in a curved gravitational
background) is $Sp(2n,\mathbb{R})$, where $n$ is the number of vector field
strengths $F^{\Lambda }$ ($\Lambda =1,...,n$). This means that the matrix%
\begin{equation}
\mathcal{S}:=\left(
\begin{array}{cc}
A & B \\
C & D%
\end{array}%
\right)  \label{S-call}
\end{equation}%
instead of being a $GL(2n,\mathbb{R})$ matrix is a symplectic matrix, indeed
it satisfies%
\begin{equation}
\mathcal{S}^{T}\Omega \mathcal{S=}\Omega ,~~~\Omega :\mathcal{=}\left(
\begin{array}{cc}
0 & -\mathbb{I} \\
\mathbb{I} & 0%
\end{array}%
\right) ,
\end{equation}%
which equivalently reads%
\begin{equation*}
A^{T}D-C^{T}B=\mathbb{I},~~A^{T}C=C^{T}A,~~B^{T}D=D^{T}B.
\end{equation*}

In absence of scalars, or if scalars are inert under duality rotations, the
e.m. duality group is at most $U(n)$ (in particular, for $n=1$, one
retrieves Maxwell e.m. duality group: $U(1)\sim SO(2)$). The constraints on
the $n$ field strength $F^{\Lambda }$ and $G^{\Lambda }$ in this case read
\begin{eqnarray}
F^{\Lambda }\star  F^{\Sigma }+ G^{\Lambda }\star G^{\Sigma } &=&0;
\label{emdeq1} \\
F^{\Lambda }\star G^{\Sigma }-F^{\Sigma }\star G^{\Lambda } &=&0.
\label{emdeq2}
\end{eqnarray}%
For $n=1$ (namely, $U(1)$), one gets%
\begin{equation}
F\star F+~G\star G=0.\label{U(1)-n=1}
\end{equation}

For general e.m. theories where also matter fields are present the
constitutive relations will also depend on these fields, that we generically
denote $\zeta ^{i}$, (and on their derivatives that we omit writing) so that
we have $G_{\mu \nu }=G_{\mu \nu }(F,\zeta ^{i})$. Duality rotations now
will include also a (nonlinear) transformation on the matter fields $\zeta
^{i}$.

As shown by Gaillard and Zumino \cite{GZ} (and reviewed in \cite{AFZ}) in
this general case, under an infinitesimal duality rotation parametrized by
\begin{equation}
s=\left(
\begin{array}{cc}
a & b \\
c & d%
\end{array}%
\right) \in \mathfrak{sp}(2n,\mathbb{R}),\mbox{ i.e. }%
~a=-d^{T},~b=b^{T},c=c^{T}~,
\end{equation}%
the Lagrangian varies by
\begin{equation*}
\delta \mathcal{L}=\frac{1}{4}\left( Fc~\star F+Gb~\star G\right) ,
\end{equation*}%
where $\delta L\equiv L(F+\delta F,\zeta ^{i}+\delta \zeta ^{i})-L(F,\zeta
^{i})$ is the total infinitesimal variation of the Lagrangian under the
duality rotations of the field strengths and of the matter fields. If we
expand $\delta F$ in terms of $F$ and $G$ using the infinitesimal version of
the duality rotation (\ref{lin-transf}) then we obtain
\begin{equation}
\delta _{\zeta }L:=L(F,\zeta ^{i}+\delta \zeta ^{i})-L(F,\chi ^{i})=\frac{1}{%
4}\left( Fc~\star F-Gb~\star G\right) +\frac{1}{2}Fd~\star G.  \label{thiseq}
\end{equation}%
If there are no matter fields the left hand side is zero, the duality
rotation group reduces then to $U(n)$, and in this case the infinitesimal
parameters satisfy $c=-b$ and $d=-d^{T}$. Then Eq. (\ref{thiseq}) reproduces
just the previous Eqs. (\ref{emdeq1}) and (\ref{emdeq2}) for $U(n)$.\newline

Note that for e.m. duality rotations the total variation $\delta \mathcal{L}$
never vanishes. If $c\not=0 , b=0$ the infinitesimal variation $\delta
\mathcal{L}$ is a total derivative because of the equations of motion $dF=0$%
. If $c=0, b\not=0$ it is a total derivative because of the equations of
motion $dG=0$. If both $b\not=0$ and $c\not=0$ we have to use both equations
of motion $dF=0$ and $dG=0$. We therefore see that the Lagrangians of
theories with duality rotation symmetry change by a term that is a total
derivative only on shell of the equations of motion $dF=0$ and $dG=0$.

We also have that under a finite duality rotation the action changes by a
term that is a total derivative only on shell of both $dF=0$ and $dG=0$
equations of motions. Indeed under a finite $U(1)$ rotation of angle $\alpha$
we have \cite{AFT}

\begin{equation}
\mathcal{L}(F^{\prime })-\mathcal{L}(F)=\frac{1}{8}\Big(sin(2\alpha
)\,(F\,\star F-G\,\star G)-4\,sin^{2}\left( \alpha \right) F\,{}\star G\Big)%
~.  \label{L'L}
\end{equation}

We notice that in the special case of Maxwell theory $G=\,\star F$ and
therefore the infinitesimal variation $\mathcal{L}(F^{\prime })-\mathcal{L}%
(F)$ is a total derivative just on shell of the relation $G=\,\star F$ and
of the equation $dF=0$ (\textit{i.e.} without using the equation of motion $%
dG=0).$ In other words, if we solve the equation $dF=0$ by introducing a
gauge potential so that $F=dA$ and hence see the Lagrangian as dependent on $%
A$, 
then for infinitesimal angle $\alpha $ the variation (\ref{L'L}) is a total
derivative \textit{\ off shell} of the equations of motion for the gauge
potential $A$. This off shell infinitesimal symmetry of Maxwell theory (that
acts non-locally on the gauge potential $A$) was studied in \cite{Deser};
however it does not correspond to a finite \textit{off shell} symmetry. As
we see from the second term in (\ref{L'L}) in order for $F\,\star G$ to be a
total derivative we have to use also the equations of motion $dG=0$, that in
Maxwell theory simply read $d\,\star F=0$.\\[1em]

Two important results are obtained from the previous formul{\ae } for
general nonlinear theories.

\begin{enumerate}
\item \textit{Any} Lagrangian density can be written as $\frac{1}{4}F{\,}%
\star G$ plus a duality invariant Lagrangian $\mathcal{L}_{inv}$,
\begin{equation}
\mathcal{L}=\frac{1}{4}F~\star G+\mathcal{L}_{inv}~,~~~\delta \mathcal{L}%
_{inv}=0.
\end{equation}%
The term $\mathcal{L}_{inv}$ provides information about the constitutive
relations (\textit{i.e.}, the medium), in particular $\mathcal{L}_{inv}=0$
vanishes in linear Maxwell theory without coupling to matter. On the other
hand, for (quadratic) theories coupled to matter $\mathcal{L}_{inv}$
provides information about some matter couplings (such as Pauli-like terms).
In supergravity it is in fact possible to construct pairs of antisymmetric
Lorentz tensors $\left( H_{\mu \nu }\left( \zeta \right) ,I_{\mu \nu }\left(
\zeta \right) \right) $ (typically bilinear in fermions) which transforms as
the couple $(F,G)$. In this case, one has%
\begin{equation}
\mathcal{L}_{inv}=\frac{1}{4}\left( FI-GH\right) +\mathcal{L}_{inv}\left(
\zeta \right) .
\end{equation}

\item The energy-momentum tensor
\begin{equation}
\Theta _{\lambda }^{\mu }=-\partial _{\lambda }\zeta ^{i}\frac{\partial
\mathcal{L}}{\partial (\partial _{\mu }\zeta ^{i})}+\delta _{\lambda }^{\mu }%
\mathcal{L}+\star {G}^{\Lambda \mid \mu \nu }F_{\Lambda |\nu \lambda }
\end{equation}%
is duality invariant if duality is a symmetry of the theory. In the absence
of matter fields the trace of the energy momentum tensor is proportional to $%
\mathcal{L}_{inv}$ : $\frac{1}{4}\Theta _{\mu }^{\mu }=\mathcal{L}_{inv}~$%
.\bigskip
\end{enumerate}

\subsection{Born-Infeld Theory}

\textit{The Born-Infeld (BI) model }\cite{BI}%
\begin{eqnarray}
\mathcal{L}_{BI} &=&\mu ^{2}\left( 1-\sqrt{-\det \left( \eta _{\mu \nu }+%
\frac{F_{\mu \nu }}{\mu }\right) }\right)  \label{BI-1} \\
&=&\mu ^{2}\left( 1-\sqrt{1+\frac{1}{2\mu ^{2}}F^{2}-\frac{1}{16\mu ^{4}}%
\left( F~\star F\right) ^{2}}\right)  \label{BI-2} \\
&=&\mu ^{2}\left( 1-\sqrt{1+\frac{1}{\mu ^{2}}\left( \left\vert
\overrightarrow{B}\right\vert ^{2}-\left\vert \overrightarrow{E}\right\vert
^{2}\right) -\frac{1}{\mu ^{4}}\left( \overrightarrow{E}\cdot
\overrightarrow{B}\right) ^{2}}\right)  \label{BI-3}
\end{eqnarray}%
\textit{\ is the simplest non-linear e.m. duality invariant theory.} It was
introduced to remove the divergence of the electrons self-energy in
classical electrodynamics. This is obtained by having an upper bound on the
electric field; indeed, for $\overrightarrow{B}=0$ $%
\mathcal{L}_{BI}$ yields the bound%
\begin{equation}
\left\vert \overrightarrow{E}\right\vert \leqslant \mu ~.
\end{equation}%
In fact, $\mathcal{L}_{BI}$ has unique properties against instabilities
created by the medium, and not many generalizations are known in the
multi-field case (\textit{i.e.}, in presence of more than one Maxwell
field). Furthermore, by expanding $\mathcal{L}_{BI}$ in power series of the
coupling constant $\frac{1}{\mu }\,$one obtains
\begin{equation}
\star G_{\mu \nu }\left( F,\star F;\mu \right) =-F_{\mu \nu }+...~
\end{equation}%
and hence for $\mu \rightarrow \infty $, \textit{i.e.}, for weak
fields, one recovers Maxwell theory.

The equations of motion of BI theory are a particular realization of the
constitutive relations
\begin{equation}
\star G_{\mu \nu }=~\star G_{\mu \nu }\left( F,\star F;\mu \right) ,
\end{equation}%
(here $\mu $ is a dimensionful coupling constant typically present in
nonlinear theories, which was previously omitted for sake of brevity). Moreover, they
satisfy the e.m. duality conditions (\ref{U(1)-n=1}).
%
%
%
%
%
%
%
By recalling (\ref{BI-1}) and the definition (\ref{orig-rels}), the explicit
dependence of $G$ on $F$ can be computed to read
\begin{equation}
G_{\mu \nu }\left( F,\star F;\mu \right) =\frac{\star F_{\mu \nu }+\frac{1}{4\mu ^{2}%
}\left( F~\star F\right) F_{\mu \nu }}{\sqrt{1+\frac{1}{2\mu ^{2}}F^{2}-%
\frac{1}{16\mu ^{4}}\left( F~\star F\right) ^{2}}}.
\end{equation}

The Hamiltonian depends on the variables $\left( \overrightarrow{D},%
\overrightarrow{B}\right) $, and is obtained by a Legendre transform with
respect to $\overrightarrow{E}$, namely by defining%
\begin{equation}
\overrightarrow{D}=\frac{\delta \mathcal{L}\left( E,B\right) }{\delta
\overrightarrow{E}}~~,~~~\mathcal{H}=\overrightarrow{D}\cdot \overrightarrow{%
E}-\mathcal{L}~.
\end{equation}%
It is worth stressing that the Hamiltonian is perfectly regular for all
values of these fields variables:%
\begin{equation}
\mathcal{H}_{BI}=\mu ^{2}\left( \,\sqrt{1+\frac{1}{\mu ^{2}}\left(
\left\vert \overrightarrow{D}\right\vert ^{2}+\left\vert \overrightarrow{B}%
\right\vert ^{2}\right) +\frac{1}{\mu ^{4}}\left\vert \overrightarrow{D}%
\wedge \overrightarrow{B}\right\vert ^{2}}-1\right) .
\end{equation}%
In particular in the limits of weakly and of strongly valued fields:%
\begin{eqnarray}
\lim_{\mu \rightarrow \infty }\mathcal{H}_{BI} &=&\frac{1}{2}\left(
\left\vert \overrightarrow{D}\right\vert ^{2}+\left\vert \overrightarrow{B}%
\right\vert ^{2}\right) ~; \\
\lim_{\mu \rightarrow 0}\mathcal{H}_{BI} &=&\left\vert \overrightarrow{D}%
\wedge \overrightarrow{B}\right\vert .
\end{eqnarray}

As yielded by the property (\ref{susyb13}), the BI Lagrangian $\mathcal{L}%
_{BI}$ is the bosonic part of the $\mathcal{N}=1$ non-linear Goldstino
Lagrangian for the aforementioned partial breaking $\mathcal{N}=2\rightarrow
1$ of rigid supersymmetry; in this context, the Maxwell field is the $%
\mathcal{N}=1$ superpartner of the spin-$\frac{1}{2}$ Goldstino field \cite%
{DP, CF,BG-1}. For various recent developments on BI theory, supersymmetry, and e.m.
duality, see \textit{e.g.} \cite{K-1,K-2,K-3,Bellucci,AnFT}.\newline

\subsection{Supersymmetry, Special Geometry, and Multi--Field Extension}\label{SUSY-SKG}

The supersymmetric BI Lagrangian follows from an algebraic constraint in Superspace among various multiplets \cite{nilpotent_2,nilpotent_3}.

This constraint enforces a non--linear relation between the two $\mathcal{N}=1$ supermultiplets $X$ and $W_\alpha$ that build the $\mathcal{N}=2$ vector multiplet \cite{BG-1,RT,Tsey}:
\beq
{W^2} \ + \ X\left( m_1\ - \ \bar{D}^2 \bar{X}\right) \ = \ 0 \ ,
\eeq{susyb18}
 where $m_1$ corresponds to a particular choice of a magnetic FI term. The non--linear relation \eqref{susyb18} determines $X$ as a non--linear function of $D^2\,W^2$ and $\overline{D}^{\,2}\,\overline{W}^{\,2}$ and implies the nilpotency constraints
\beq
X^2 \ = \ 0 \ , \quad  X \, W_\alpha \ = \ 0 \ .
\eeq{susyb1856}
The supersymmetric BI Lagrangian then acquires the form
\beq
{\cal L}_{SBI} \ = \ Im \, \int d^2 \theta \ (e_1+ie_2) \ X\left(m_1, D^2\,W^2,\overline{D}^{\,2}\,\overline{W}^{\,2}\right) \ ,
\eeq{susyb19}
where $e_1$ and $e_2$ are electric FI terms whose labels correspond to the first two directions in an $SU(2)$ triplet. For canonically normalized vectors, one can then recognize that the scale $\mu$ and the theta--angle $\vartheta$ can be expressed in terms of $m_1$, $e_1$ and $e_2$ as
\beq
\mu \ = \ \sqrt{m_1\,e_2} \ , \quad \vartheta \ = \ \frac{e_1}{e_2} \ .
\eeq{susyb20}
In the bosonic limit, the quadratic constraint becomes
\begin{equation}
F_{+}^{2}+\mathcal{F}\left( m_1-\mathcal{\bar{F}}\right) =0,  \label{qc}
\end{equation}%
where $F_{+}$ is the \textit{self-dual curvature} of the Maxwell field :%
\begin{equation}
F_{+\mid \mu \nu }:=\frac{1}{2}\left( F_{\mu \nu }+i~\star F_{\mu \nu
}\right)
\end{equation}%
and $\mathcal{F}$ is an auxiliary complex scalar field such that%
\begin{eqnarray}
\mathcal{L}_{BI}&&=e_1\ \text{Im}\left( \mathcal{F}\right)+e_2\ \text{Re}\left( \mathcal{F}\right)= \nonumber\\
&&=-\vartheta F_{\mu\nu}\,\star {F}^{\mu\nu} +\frac{\mu^2}{2}\left[ \,1 \ - \ \sqrt{1 \ + \ \frac{4}{\mu^2} \ F_{\mu\nu}\,F^{\mu\nu} \ - \ \frac{4}{\mu^4} \ \left( F_{\mu\nu}\,\star {F}^{\mu\nu} \right)^2}\,\right].\label{n1}
\end{eqnarray}

In the multi-field case, the generalization of $\mathcal{L}_{SBI}$ rests on
the constraints ($A=1,...,n$) \cite{fps,fs}
\beq
d_{ABC}\left[ W^B, W^C + Y^B\left( m^C - \bar{D}^2 \bar{Y}^C\right) \right] \ = \ 0 \ ,
\eeq{30ab}
and the complete Lagrangian reads
\beq
{\cal L} \ = \ - \ Im \, \int d^2 \theta \, \left[U_{AB}\,W^A\,W^B \ + \ {\cal W}(X) \ + \ \frac{1}{2} \  \bar{D}^2 \, \left(X^A\, \bar{U}_A\ -\ \bar{X}^A,  U_A\right)\right] \ .
\eeq{29}
The $\theta^2$ component of eq.~\eqref{30ab},
\beq
d_{ABC} \left[ F_{+}^B \cdot F_{+}^C \ + \  {\cal F}^B \left( m^C \, -\, \bar{\cal F}^C \right)  \right] \ = \ 0 \ ,
\eeq{30d}
where $F_{+}^A$ are self--dual field strength combinations and ${\cal F}^A$ is the auxiliary field,  is the multi--field generalization of the BI constraint \eqref{qc} that is induced by the Special Geometry. Moreover, it holds that
\beq
\Re \, F_+^A\ F_+^B \ =  F^A\ F^B \ , \qquad \Im \, F_+^A\ F_+^B \ =  F^A\ \star F^B \ .
\eeq{3}
$d_{ABC}$ is a rank-$3$ completely symmetric tensor,
encoding the coefficients of the cubic term of the holomorphic prepotential
of the aforementioned rigid special K\"{a}hler geometry characterizing the $%
\mathcal{N}=2$ vector multiplets' scalar manifold, whose classification is
based on the singularity structure of cubic varieties \cite{fs,fpssy}.

The real parts of Eqs.~\eqref{30d} are $n$ quadratic equations that are generally coupled. Letting
\beq
H^A \ = \ \frac{m^A}{2} \ - \ \Re \mathcal{F}^A \ , \qquad R^{AB} \ = \ F^A\ F^B \ + \frac{m^A\,m^B}{4} \ - \ \Im \mathcal{F}^B \, \Im \mathcal{F}^C \ ,
\eeq{4}
they take the form
\beq
d_{ABC} \left( H^B\, H^C \ - \ R^{BC} \right) \ = \ 0 \ .
\eeq{5}
On the other hand, the imaginary parts of Eqs.~\eqref{30d} are $n$ linear equations for $\Im \mathcal{F}^A$:
\beq
d_{ABC} \left( F^B \ {\star F}^C \ + \ \Im \mathcal{F}^B \, m^C \right) \ = \ 0 \ .
\eeq{6}

The bosonic part of the multi-field Lagrangian (\ref{29}) can be expressed in terms of the $\mathcal{F}^A$ and of the real magnetic changes $m^A$. It can also be written in terms of additional complex charges $e_A \ = \ e_{1\,A}+i\ e_{2\,A}$, as
\beq
{\cal L}_{\rm Bose} \ = \ e_{2A} \left( \frac{m^A}{2} \ - \ H^A \right) \ + \ C_{AB} \left( H^A \, H^B \ - \ R^{AB} \right) \ + \e_{1A}\, \Im \mathcal{F}^A \ .
\eeq{7}

The matrix $C_{AB}$ is needed whenever the matrix $d_{AB} \ :=  \ d_{ABC} \ q^C$ is not positive definite. Moreover, by a change of symplectic basis one could also eliminate the real parts $e_{1A}$ of the electric charges, which multiply combinations of the fields that are total derivatives. These Lagrangians combine, in general, a quadratic Maxwell--like term with additional higher--order contributions. For $n=1$, or whenever the matrix $d_{AB}$ is positive definite, one is not compelled to introduce the $C_{AB}$ and the Lagrangian takes the simpler form ~\eqref{n1}. In all cases, however, the difficult step in the construction of the Lagrangians is the solution of the quadratic constraints, and in particular of the non--linear ones given in~\eqref{5}.

\subsection{Schr\"{o}dinger Formulation}

The general study and classification of constitutive relations admitting
e.m. duality symmetry can be attacked by adopting complex and chiral
variables, indeed these variables simply transform with a phase under
duality symmetry rotations. We therefore define\footnote{%
Here and below, the bar denotes complex conjugation.}:
\begin{equation}
T:=F-iG\Leftrightarrow \overline{T}:=F+iG,
\end{equation}%
so that the e.m. duality condition (\ref{U(1)-n=1}) equivalently
reads
\begin{equation}
\star \overline{T}T=0.  \label{emcondT}
\end{equation}%
The corresponding chiral combinations read%
\begin{equation}
T^{\pm }:=\frac{1}{2}\left( F^{\pm }-iG^{\pm }\right) \Leftrightarrow
\overline{T^{\pm }}:=\frac{1}{2}\left( F^{\mp }+iG^{\mp }\right) ,
\end{equation}%
where $F^{\pm }=\frac{1}{2}(F\pm i\,\star F)$, $G^{\pm }=\frac{1}{2}(G\pm
i\,\star G)$. The equations of motion of linear electromagnetism in these
variables read
\begin{equation}
T^{+}=0\Leftrightarrow \left( F-iG\right) +i\left( \star F-i~\star G\right)
=0\Leftrightarrow G=~\star F.
\end{equation}%
The constitutive relations in these new variables, studied in \cite{ZI}, and
independently and more generally in \cite{BN, Kallosh:2011dp}, read as
follows:%
\begin{equation}
T^{+}=\frac{\partial I\left( T^{-},\overline{T^{-}}\right) }{\partial
\overline{T^{-}}},~~~\overline{T^{+}}=\frac{\partial I\left( T^{-},\overline{%
T^{-}}\right) }{\partial T^{-}}.  \label{4.18}
\end{equation}%
These are six relations if the arbitrary function $I$ (the \textquotedblleft
Lagrangian\textquotedblright\ of these new variables) is real. Moreover,
duality invariance is obtained if $I$ is invariant under duality rotations.
This is the case if
\begin{equation*}
I\left( T^{-},\overline{T^{-}},g\right) =I(u),
\end{equation*}%
where
\begin{equation}
u=2g^{2}|(T^{-})^{2}|=g(|T^{2}|+|T\,^{\ast }T|)~,~~\mbox{ i.e., }%
~~u^{2}=4g^{4}(T^{-})^{2}(\overline{T}^{-})^{2}~.
\end{equation}%
%
%
%
%
%
%
We note that the function $I$ is $U(1)$-duality-invariant because the
Lorentz scalar $(T^{-})^{2}(\overline{T^{-}})^{2}$ is duality invariant,
indeed under a $U(1)$ duality rotation of angle $\alpha $, the fields $T$
and $\overline{T}$ transform with opposite phases $e^{-i\alpha }$ and $%
e^{i\alpha }$.\newline


By use of a Legendre transform \cite{ZI, AF, AFT} between the functions ${%
\mathcal{L}}(F)$ and $I(T^{-},\overline{T^{-}})$, the constitutive relations
(\ref{4.18}) in the $T$-variables can be shown to be equivalent to the
original constitutive relations (\ref{orig-rels}).\newline
\newline
To summarize, the freedom in writing a self-dual and non-linear theory of
electromagnetism can be traced back to a real function $I$ of a real
(Lorentz-invariant and $U(1)$-duality invariant) variable $\left(
T^{-}\right) ^{2}\left( \overline{T^{-}}\right) ^{2}$, or its square root $u$%
. Arbitrary functions $I(u)$ are in one-to-one correspondence with
Lagrangians $\mathcal{L}\left( F\right) $ such that the equations (\ref{4.18}%
) are equivalent to the original constitutive relations and self duality
condition (\ref{orig-rels}) and (\ref{U(1)-n=1}).

Furthermore, it was shown by Schr\"{o}dinger \cite{Schr} (see also \cite{GZ2}%
) that the constitutive relations for the BI theory are
\begin{equation}
\star T_{\mu \nu }=-\frac{T^{2}}{\star TT}T_{\mu \nu }-\frac{g^{2}}{8}~\star
TT\,\overline{T}_{\mu \nu }.
\end{equation}%
In the general case the following constitutive relations \textit{\`{a} la
Schr\"{o}dinger} define a nonlinear and self-dual theory \cite{AF}
\begin{equation}
\star T_{\mu \nu }=-\frac{T^{2}}{\star TT}T_{\mu \nu }-\frac{g^{2}}{8}\frac{%
f(u)}{u}~\star TT\,\overline{T}_{\mu \nu },  \label{generalTemd}
\end{equation}%
where we recall that $u=2|(T^{-})^{2}|=|T^{2}|+|T\,\star T|\,,$ and where we
require $lim_{u\rightarrow 0}f(u)=0$, so to recover Maxwell theory for weak
fields. For example the duality conditions (\ref{emcondT}) follow
contracting (\ref{generalTemd}) with $\star T^{\mu \nu }$. Contracting (\ref%
{generalTemd}) with $T^{\mu \nu }$ and with $\overline{T}^{\mu \nu }$ it can
also be shown that
\begin{equation}
\left( 1-t^{2}\right) ^{2}f(u)=32t;
\end{equation}%
where
\begin{equation}
t=\frac{T\overline{T}}{|T^{2}|+|T\,\star T|}~.
\end{equation}

In the first Ref. of \cite{Kallosh:2011dp} the function $t(u)$ corresponding
to the BI Lagrangian density was found through an iterative procedure order
by order in $u$; the first coefficients of its power series expansion were
observed to match those of a particular generalized hypergeometric
function, yielding to the conclusion that%
\begin{eqnarray}
t(u) &=&\frac{u}{32}~_{3}F_{2}\left( \frac{1}{2},\frac{3}{4},\frac{5}{4};%
\frac{4}{3},\frac{5}{3};-\frac{u^{2}}{3^{3}\cdot 2^{2}}\right) \\
&=&\frac{u}{16}\sum_{k=0}^{\infty }(-1)^{k}\frac{(4k+1)!}{(3k+2)!k!}\left(
\frac{u^{2}}{4^{5}}\right) ^{k}.  \label{426}
\end{eqnarray}%
The relation between the constitutive relations \textit{\`{a} la Schr\"{o}%
dinger} and those in terms of the $I(u)$ function follows from the relations
between the $T$ variables and the chiral $T^{\pm }$ variables. This allows
to determine $f(u)$ from $I(u)$ and vice versa. Explicitly, given $f(u)$,
one obtains $t=t(u)$ from (\ref{426}); the function $I(u)$ is then given by
\begin{equation}
I\left( u\right) =\frac{1}{2}\int t(u)du.
\end{equation}

\subsection{Generalizations}

Other non-linear (square-root) BI-like type Lagrangian densities in $D$
dimensions have been suggested in \cite{FS,FSY}, starting from alternative $%
D=4$ Goldstino Lagrangians studied in \cite{BG-2,RT,KT}.

The first class of Lagrangians is given by $D$-dimensional generalizations
containing pairs field strengths forms of degrees $p+1$ and $D-p-1$,
corresponding to gauge fields coupling to $\left( p-1\right) $-- and $\left(
D-p-3\right) $-- branes, respectively. These Lagrangians generalize the $D=4$%
, $p=2$ case, corresponding to a non-linear Lagrangian for a tensor
multiplet regarded as an $\mathcal{N}=2\rightarrow 1$ Goldstino multiplet in
rigid supersymmetry \cite{KT,FSY}. They read%
\begin{eqnarray}
\mathcal{L}\left( X,Y^{2};\mu \right) &=&\mu ^{2}\left( 1-\sqrt{1+\frac{1}{%
\mu ^{2}}X-\frac{1}{\mu ^{4}}Y^{2}}\right) ; \\
X &:&=-\star \left( H_{p+1}\wedge \star H_{p+1}+V_{D-p-1}\wedge \star
V_{D-p-1}\right) ; \\
Y &:&=\star \left( H_{p+1}\wedge V_{D-p-1}\right) ,
\end{eqnarray}%
and they have the property of being \textit{doubly self-dual} under%
\begin{equation}
V_{D-p-1}^{\prime }=\star H_{p+1};~~~H_{p+1}^{\prime }=\star V_{D-p-1}.
\end{equation}%
Moreover, after a single duality the Lagrangian ends up containing two forms
of the same degree, and it can be recast in the following manifestly $%
U(1)_{e.m.}$-invariant form \cite{FSY}:%
\begin{equation}
\mathcal{L}^{\prime }\left( W,\bar{W};\mu \right) =\mu ^{2}\left[ 1-\sqrt{1+%
\frac{\left(W_{D-p-1}\ \bar{W}_{D-p-1}\right)}{\mu ^{2}}+\frac{\left( W_{D-p-1}\
\bar{W}_{D-p-1}\right) ^{2}-W_{D-p-1}^{2}\bar{W}_{D-p-1}^{2}}{4\mu ^{4}}}%
\right] .
\end{equation}%
Actions with the full $U(n)$ (where $n$ is the number of Maxwell vector
fields) were proposed in \cite{ABMZ}, but they are currently not available
in closed form, even for $n=2$.\medskip

On the other hand,  we can present three relatively simple extensions of the BI Lagrangian to a pair of curvatures $F^1$ and $F^2$ whose field equations possess $U(1)$, $U(1)\times U(1)$ and $SU(2)$ duality symmetries. All these models solve subsets of the non-linear Gaillard-Zumino
constraints (\ref{emdeq1}) and (\ref{emdeq2}) in the case $\Lambda ,\Sigma =1,2$, \textit{i.e.} of :%
\begin{eqnarray}
F^{1}\star F^{2}+G^{1}\star G^{2} &=&0;\label{unno} \\
F^{1}\star F^{1}+G^{1}\star G^{1} &=&0; \\
F^{2}\star F^{2}+G^{2}\star G^{2} &=&0; \\
F^{1}\star G^{2}-F^{2}\star G^{1} &=&0.
\end{eqnarray}

The Lagrangian with $U(1)$ invariance reads
\beq
{\cal L} \ = \mu^2 \left[ \, 1\ -\ \sqrt{1\,+\,\frac{(F^1)^2+(F^2)^2}{\mu^2}
\,-\,\frac{\left(\star\left[F^1 \wedge F^2  \right]\right)^2}{\mu^4}}\,\right] \ ,
\eeq{B65}
and was obtained as an application of double duality between two forms of the same degree. This $U(1)$ corresponds to the constraint (85) for $\Lambda =1$ and $\Sigma=2$, namely (\ref{unno}).

The other two examples are simply different complexifications of the two--field system, whose manifest $U(1)$ invariance corresponds to the single constraint of Eq.~\eqref{emdeq2}. In particular, letting
\beq
F \ = \ F^1 \ + \ i\, F^2 \ ,
\eeq{B66}
the case with $U(1) \times U(1)$ invariance, where the second $U(1)$ constraint corresponds to the equation
\begin{equation}
F^1\star F^1+F^2\star F^2+G^1\star G^1+G^2\star G^2=0,
\end{equation}
reads
\beq
{\cal L} \ = \mu^2 \left[ \, 1\ -\ \sqrt{1\,+\,\frac{F\ \overline{F}}{\mu^2}
\,-\,\frac{\left(\star \left[F \wedge \overline{F} \right]\right)\,\left(\star \left[{F} \wedge \overline{F} \right]\right)}{\mu^4}}\,\right] \ .
\eeq{B67}
This is actually a particular case of the $U(n,n)$ dualities of \cite{ABMZ} for $n=1$, where in the absence of scalars $U(1,1)$ reduces to its maximal compact subgroup $U(1) \times U(1)$. Finally, the case with $SU(2)$ invariance reads
\beq
{\cal L} \ = \mu^2 \left[ \, 1\ -\ \sqrt{1\,+\,\frac{F\ \overline{F}}{\mu^2}
\,-\,\frac{\left(\star \left[F \wedge F \right]\right)\,\left(\star \left[\overline{F} \wedge \overline{F} \right]\right)}{\mu^4}}\,\right] \ .
\eeq{B13}
where the other two constraints correspond to the $SU(2)$ generators
\begin{eqnarray}
&&F^1\star F^1-F^2\star F^2+G^1\star G^1-G^2\star G^2=0,\\
&&F^1\star F^2+G^1\star G^2=0.
\end{eqnarray}
From these examples it is manifest that the standard BI system admits several types of inequivalent complexifications that differ in their quartic couplings.

A fourth option, with maximal $U(2)$ duality, is naturally captured by the construction of \cite{ABMZ}, but its Lagrangian, as have been mentioned before, is not known in closed form.

\medskip

All the above non-linear actions can be \textit{massive} by introducing
Green-Schwarz terms \cite{FS}, \textit{i.e.} couplings to another gauge
field of the form%
\begin{equation}
mH_{p+1}\wedge A_{D-p-1}.
\end{equation}%
A St\"{u}ckelberg mechanism is then generated, and the non-linear mass terms
turn out to take the same functional form as the original non-linear
curvature terms; in other words, the $\left( D-p-1\right) $-form gauge field
has been eaten to give mass to the original $p$-form gauge field (whose
field strength is $H_{p+1}$). The simplest example of this class of
Lagrangians is the $D=4$ BI action used to make an antisymmetric field $B_{\mu
\nu }$ massive :
starting point is thus the master action
\bea
{\cal L} &=& \ - \frac{k^2}{12} \ H_{\mu\nu\rho}\, H^{\mu\nu\rho} \ - \ \frac{m}{4} \ \epsilon^{\mu\nu\rho\sigma} B_{\mu\nu}\, F_{\rho\sigma} \nonumber \\ &+& \frac{\mu^2}{8\, g^{\,2}} \ \left[ \,1 \ - \ \sqrt{1 \ + \ \frac{4}{\mu^2} \ F_{\mu\nu}\,F^{\mu\nu} \ - \ \frac{4}{\mu^4} \ \left( F_{\mu\nu}\,\star {F}^{\mu\nu} \right)^2}\,\right]\ ,
\eea{13}
where we have introduced a dimensionless parameter $g$, the counterpart of the parameter $k$ that accompanies the two--form kinetic term. The parameter $\mu$ is the BI scale factor, with mass--squared dimension, which sizes the non--linear corrections.

As above, a massive variant of the BI action principle would be obtained by eliminating $H$ after moving to a first--order form where it is unconstrained. However, as we have seen, the additional field is just a standard St\"{u}ckelberg mode, so that for brevity we can just display the gauge--fixed Proca--like Lagrangian for the massive BI vector,
\beq
{\cal L} \ = \ \ - \frac{m^2}{2\, k^2} \ A_{\mu}\, A^{\mu} \ + \ {\cal L}_{BI} \left(g,\mu, F_{\mu\nu} \right) \ ,
\eeq{14}
where
\beq
{\cal L}_{BI}\left(g,\mu, F \right) \ = \ \frac{\mu^2}{8\, g^{\,2}} \ \left[ \,1 \ - \ \sqrt{1 \ + \ \frac{4}{\mu^2} \ F_{\mu\nu}\,F^{\mu\nu} \ - \ \frac{4}{\mu^4} \ \left( F_{\mu\nu}\,\star {F}^{\mu\nu} \right)^2}\,\right] \ .
\eeq{14a}

In the massless case, the self--duality of the BI Lagrangian would translate into the condition that
\beq
{\cal L}_{BI} \bigg(g\, ,\, \mu\, ,\, F_{\mu\nu}(A) \bigg)  \ = \ {\cal L}_{BI} \left(g^\prime=\frac{1}{g}\, , \, \mu^\prime = \frac{\mu}{g^{\,2}}\, , \, F_{\mu\nu}(C) \right) \ ,
\eeq{14b}
where $C$ is the dual gauge field. On the other hand, in the presence of the Green--Schwarz term ($m \neq 0$), one can eliminate the vector altogether and work, in the dual formulation, solely in terms of the two--form $B_{\mu\nu}$. The self--duality of the massless BI theory then implies that the dual action involves a BI mass term and reads
\bea
{\cal L} &=& - \ \frac{k^2}{12} \ H_{\mu\nu\rho}(B)\, H^{\mu\nu\rho}(B) \nonumber \\ &+& \frac{\mu^2}{8\,g^{\,2}} \ \left[ \,1 \ - \ \sqrt{1 \ + \ \frac{4\, m^2\,g^{\,4}}{\mu^2} \ B_{\mu\nu}\,B^{\mu\nu} \ - \ \frac{4\, m^4\, g^{\, 8}}{\mu^4} \ \left( B_{\mu\nu}\,\star {B}^{\mu\nu} \right)^2}\,\right] \ .
\eea{15}
The massless limit can be recovered reintroducing the gauge invariant combination $m \, B  \ + \ dC$ before letting $m \to 0$. In this fashion, the limiting Lagrangian describes a massless two--form, dual to a scalar, and a dual massless vector $C$.

This procedure could be relevant when
the system is coupled to $\mathcal{N}=2$ supergravity, in which case one of
the two gravitino would belong to a massive spin-$\frac{3}{2}$ multiplet.

\vskip 24pt

\noindent {\large \textbf{Acknowledgements}}\newline
\noindent We are grateful to L. Andrianopoli, P. Aschieri, R. D'Auria, M.
Porrati, R. Stora, S. Theisen, M. Trigiante, and especially to A. Sagnotti for useful
discussions and/or collaborations.

\end{document}